\documentclass[pra,reprint,floatfix,amsmath,amssymb,aps]{revtex4-2}
\usepackage{graphicx}% Include figure files
\usepackage{dcolumn}% Align table columns on decimal point
\usepackage{bm}% bold math
\usepackage{hyperref}% add hypertext capabilities
\usepackage{braket}
\hypersetup{
    colorlinks,
    linkcolor={red},
    citecolor={blue},
    urlcolor={blue}
}

\def\nn{\nonumber \\ }
\newcommand{\be}{\begin{equation}}
\newcommand{\ee}{\end{equation}}
\newcommand{\bea}{\begin{eqnarray}}
\newcommand{\eea}{\end{eqnarray}}

\def\fr#1{(\ref{#1})}

\DeclareMathOperator{\Tr}{Tr}
\DeclareMathOperator{\sgn}{sgn}
\newcommand{\Z}{\mathbb{Z}}
\renewcommand{\Re}{\operatorname{Re}}
\renewcommand{\Im}{\operatorname{Im}}
\newcommand{\odiff}[2]{\frac{{\rm d}#1}{{\rm d}#2}}

\renewcommand{\vec}[1]{\boldsymbol{#1}}

\begin{document}

\preprint{APS/123-QED}

\title{Decay of long-lived oscillations after quantum quenches in
gapped interacting quantum systems} 

\author{Jacob H. Robertson}
 \email{jacob.robertson@physics.ox.ac.uk}
\author{Riccardo Senese}%
\author{Fabian H. L. Essler}%
\affiliation{%
    Rudolf Peierls Centre for Theoretical Physics, Oxford University, Oxford OX1 3PU, United Kingdom
}%

\date{\today} %Update this on resubmission

\begin{abstract}
The presence of long-lived oscillations in the expectation values of
local observables after quantum quenches has recently attracted considerable
attention in relation to weak ergodicity breaking. Here we focus on an 
alternative mechanism that gives rise to such oscillations in a class of systems
that support kinematically protected gapped excitations at zero temperature.
An open question in this context is whether such
oscillations will ultimately decay. We provide strong support for the decay hypothesis by considering 
spin models that can be mapped to systems of weakly interacting fermions, 
which in turn are amenable to an analysis by standard methods based on the 
BBGKY hierarchy. We find that there is a time scale 
beyond which the oscillations start to decay which grows as the strength of the quench is made small.
\end{abstract}

\maketitle

\section{\label{sec:Intro} Introduction}

A key question in non-equilibrium many-body quantum dynamics is to 
understand how ergodicity can be broken and thermalization avoided. The 
best known examples are many-body localization 
(MBL)\cite{Abanin2019Colloquium} and quantum integrable
systems \cite{*Rigol2007Relaxation,EsslerQuench2016}. 
The issue of ergodicity in generic systems is addressed by the Eigenstate 
Thermalization Hypothesis \cite{DeutschETH1991,
SrednickiETH1994,DAlessioETH2016} (ETH), which provides a description of the structure 
of the matrix elements of observables in energy eigenstates.
In the literature, violations of ergodicity are
often referred to as either `strong' or `weak' depending on whether
the fraction of states failing to satisfy ETH remains non-zero or vanishes
in the thermodynamic limit, respectively \cite{Kim2014Testing}. Weak violations of
ergodicity have recently attracted a great deal of attention in the
context of so-called quantum many-body scars (QMBS)
\cite{Turner2018Weak,moudgalya2018entanglement,Serbyn2021Quantum,Moudgalya2022Quantum,Chandran2022Quantum}.
This notion was first introduced to explain the unexpected dynamics of a
Rydberg atom quantum simulator \cite{BernienProbing2017} where 
initializing the system in a particular initial state resulted in
long-lived oscillations in the time-evolution of observables. 

A seemingly related phenomenon has been observed in quenches in
certain spin chains, referred to as `weak
thermalization' in \cite{Banuls2011Strong}. Various authors
\cite{Banuls2011Strong,Rakovszky2016Hamiltonian,KormosRealtime2016,Lin2017Quasiparticle,Collura2018Dynamical,Hodsagi2018Quench,Robinson2019Signatures,Castro_Alvaredo2020Entanglement,Scopa_2022,Vovrosh_2022,Birnkammer_2022}
have noted that for the Ising chain in a tilted field following a
quench, there are long-lived oscillations at frequencies corresponding
to the masses of `meson' bound states. Similar oscillations were observed in the Potts model \cite{Pomponio2019Quasi}.
    These studies have been unable to simulate late enough times to conclude if the meson oscillations decayed at late times. In the paramagnetic regime \cite{Rakovszky2016Hamiltonian} found oscillations beginning to damp following a quantum quench in the scaling limit.
Theoretical arguments for the eventual decay have been given in \cite{Lin2017Quasiparticle}. In the latter work it is moreover argued
that such oscillations should be generic to quantum systems with a
quasiparticle gap and isolated bands such as produced by bound
states. 

The fate of oscillations at late times is however controversial:
oscillations in the post-quench dynamics of observables in quantum
field theories were predicted by Delfino and collaborators in \cite{Delfino2014Quantum,Delfino2017Theory} and subsequently in \cite{Delfino2020Persistent,Delfino2022persistent}
it was argued that they persist for arbitrarily long
times.

The Ising chain in a tilted field is unusual in the sense that the
perturbation that leads to the formation of meson bound states is
non-local with respect to the fermionic elementary excitations of the
transverse-field Ising chain. This precludes the analysis of particle
decay by standard perturbative approaches. In light of this fact it is
important to identify models that exhibit the same phenomenology, but
can be studied by such methods. One such example was recently reported
by us in the axial next-nearest neighbour Ising model
(ANNNI)~\cite{Robertson2023Simple}, which has a non-confining (contact)
potential for the elementary fermion excitations.
In this work we give two quench setups that we consider to be particularly simple examples of such behaviour. The first is a quantum quench in integer spin antiferromagnetic chains, which possess a gapped single particle (magnon) mode according to Haldane's conjecture~\cite{Haldane1983Continuum,Haldane1983Nonlinear}. This model very cleanly demonstrates that the phenomenology is due neither to confinement, nor indeed to bound states, but simply due to the system possessing a kinematically protected gapped quasiparticle excitation. However, as integer spin antiferromagnets are strongly interacting systems
we are restricted to purely numerical investigations of the non-equilibrium dynamics of this model by matrix product state methods.
Secondly, we present and analyze in detail a novel example of these long-lived oscillations in a model with weak interactions that has the simplifying feature of exhibiting a $\mathsf{U}(1)$ symmetry associated with particle number conservation: a dimerized XXZ chain in a staggered magnetic field. In the scaling limit, the low-lying excitations of this model can be understood in terms of solitons, anti-solitons and a bound state known as a `breather' which can give rise to long-lived oscillations after quantum quenches. 

Many other mechanisms for producing long-lived oscillations are possible in quantum systems that should be differentiated from the case we discuss.
We have already mentioned exact quantum many-body scars, which can cause  infinite lifetime oscillations if the initial state has large overlap with scar states lying in the middle of the spectrum. 
There are also Bloch oscillations of domains in the tilted field Ising model at low domain-wall density and related models of Rydberg atoms \cite{Balducci2023Interface, Magoni2021Emergent}.  
Long-lived oscillations can also occur in the electric field strength in lattice gauge theories that are related to quantum many-body scars \cite{Surace2020Lattice,Lerose2020Quasilocalized}. 
Finally, in the presence of a $\mathsf{U}(1)$ symmetry undamped oscillations can occur when one considers observables that connect neighbouring charge sectors and applies a Zeeman field that splits all sectors by the same energy difference \cite{Vorndamme2021Observation,reimann2023nonequilibration}.

The organization of this paper is as follows: in Section~\ref{Sec:XXZ} we describe the mechanism that can give rise to long-lived oscillations in models with kinematically protected single-particle excitations before giving a simple numerical case study of such behaviour in the spin-1 bilinear biquadratic (BLBQ) chain in Section~\ref{subsec:spin1}.
The spin-1 chain provides a clear example of the phenomenology we are interested in, however it is always strongly interacting.
In contrast, the ANNNI and related Ising models can be mapped to fermionic chains for which the interaction can be considered perturbatively. However the lack of a $\mathsf{U}(1)$ symmetry (and in the case of a tilted field, long-ranged interactions between fermions) significantly complicates the application of standard methods based on the BBGKY hierarchy \cite{bogolûbov1970lectures,huang2008statistical,bonitz2015quantum}. In Section~\ref{subsec:XXZ} we address this problem by introducing a dimerized XXZ chain in a staggered magnetic field, which exhibits a $\mathsf{U}(1)$ symmetry as well as long-lived oscillations of observables after quantum quenches.
In Section~\ref{Sec:Decay} we present two different approximations based on the BBGKY hierarchy --- a self-consistent time-dependent mean-field theory (SCTDMFT) and the Second Born approximation \cite{Bertini15Prethermalization,Bertini16Thermalization} --- to study the quench dynamics of local observables. 

%%%%%%%%%%%%%%%%%%%%%%%%%%%%%%%%%%%%%%%%%%%%%%%%%%%%%%%%%%%%%%%%%%%
\section{\label{Sec:XXZ} Oscillations at ``early" times}
%%%%%%%%%%%%%%%%%%%%%%%%%%%%%%%%%%%%%%%%%%%%%%%%%%%%%%%%%%%%%%%%%%%
The physics underlying the oscillations explored within this paper arises from two key requirements, these are:
\begin{enumerate}
\item{} The existence of a kinematically protected mode, i.e. a quasiparticle with a gapped dispersion $\epsilon(q) \geq \Delta_{\rm ex}$ such that there is some region in the energy-momentum plane that has no other energy eigenstates. 
\item{} Initial density matrices $\rho(t=0)$ such that the energy density deposited into the system by the quench (relative to the post-quench ground state energy $E_{\rm GS}$) is small compared to the spectral gap of the post-quench Hamiltonian
\be
\epsilon_{\rm Quench}=\lim_{L\to\infty}\frac{1}{L}\left({\rm Tr}\big[\rho(t=0) H\big]-E_{\rm GS}\right)\ll
\Delta_{\rm ex}.
\ee
\end{enumerate}
If the above requirements are met the physics can be viewed in terms of a dilute gas of long-lived particles, whose scattering is to a good approximation purely elastic, \emph{cf.}
\cite{james2008finite,goetze2010low,essler2008finite,pozsgay2008form,essler2009finite,pozsgay2010form,CEF1,granet2020finite}. The possible emergence of long-lived oscillations can then be 
understood by considering the linear response regime of ground state quenches. This is equivalent to the approach of Refs~\cite{Delfino2014Quantum,Delfino2017Theory,Delfino2020Persistent,Delfino2022persistent}. To that end we consider an initial state $|\psi_0\rangle$ that is the ground state of a Hamiltonian $H_0$, and time-evolve with $H=H_0+\lambda V$. Here $V$ is a global operator that is assumed to be translationally invariant (as is $H_0$). By construction both $H_0$ and $H$ feature kinematically protected gapped single-particle excitations.
Linear response theory then gives
\begin{align}
\langle\psi_0|{\cal O}(t)|\psi_0\rangle&\approx
\langle\psi_0|{\cal O}|\psi_0\rangle
-i\lambda\int_0^t dt'\ \chi_{{\cal O}V}(t,t')\ ,\nn
\chi_{{\cal O}V}(t,t')&=
\langle\psi_0|[{\cal O}_I(t),V_I(t')]|\psi_0\rangle,\nn
{\cal O}_I(t)&\equiv e^{iH_0t}{\cal O}e^{-iH_0t}\ .
\end{align}
The response function $\chi_{{\cal O}V}(t,t')$ can be expressed in terms of a Lehmann representation using the eigenstates of $H_0$, which gives
\be
\chi_{{\cal O}V}(t,t')=\sum_n e^{i(t-t')(E_0-E_n)}\langle\psi_0|{\cal O}|\psi_n\rangle
\langle \psi_n| V|\psi_0\rangle-{\rm c.c.}
\ee
We now consider perturbations $V$ and operators ${\cal O}$ that have non-vanishing matrix elements between the ground state and the kinematically protected gapped excitation. As by construction we are dealing with a ground state calculation the contribution of this excited state will provide the dominant contribution to the linear response function for large $t$
\begin{align}
&\langle\psi_0|{\cal O}(t)|\psi_0\rangle\approx
\langle\psi_0|{\cal O}|\psi_0\rangle\nn
&+2\lambda\sum_k {\rm Re}\frac{e^{-it\bar{\epsilon}(k)}-1}{\bar{\epsilon}(k)}
F_{\cal O}(k)F^*_{V}(k)
+\dots
\label{linearresponse}
\end{align}
where $F_{\cal O}(k)=\langle\psi_0|{\cal O}|k\rangle$ is the matrix element of the operator ${\cal O}$ between the ground state of $H_0$ and the single quasiparticle excitation of $H_0$ with momentum $k$ and dispersion $\bar{\epsilon}(k)$. 
As $V$ is by assumption a global, translationally invariant operator the only non-zero matrix element is with the zero-momentum single-particle state
\be
F_V(k)\propto \delta_{k,0}\ ,
\ee
which establishes that in linear response theory we obtain persistent oscillations with frequency $\bar{\epsilon}(0)$, \emph{cf.} Refs~\cite{Delfino2014Quantum,Delfino2017Theory,Delfino2020Persistent,Delfino2022persistent}. If instead either $V$ or the initial state has invariance only under translations by two sites, the matrix element will select out momenta $k=0$ and $k=\pi$. In this case oscillations will occur at two frequencies, $\bar{\epsilon}(0), \bar{\epsilon}(\pi)$, as long as the kinematically protected mode exists at these momenta.
The approach of \cite{Delfino2014Quantum,Delfino2017Theory,Delfino2020Persistent,Delfino2022persistent} can be straightforwardly modified by expanding the initial state in terms of the eigenstates of the post-quench Hamiltonian using perturbation theory. This gives
\begin{align}
&\langle \psi_0|{\cal O}(t)|\psi_0\rangle \approx
\langle \psi_0|{\cal O}|\psi_0 \rangle \nn
&+ 2\lambda\sum_k {\rm Re}\frac{e^{-it{\epsilon}(k)}-1}{{\epsilon}(k)}
\widetilde{F}_{\cal O}(k)\widetilde{F}^*_{V}(k)
+\dots
\end{align}
Here ${\epsilon}(k)$ is the dispersion of the kinematically protected mode in $H$ (rather than $H_0$) and
$\widetilde{F}_{\cal O}(k)=\langle 0|{\cal O}|\widetilde{k}\rangle$ is the matrix element of the operator ${\cal O}$ between the ground state of $H$ and the single quasiparticle excitation of $H$ with momentum $k$. This way of approaching the problem is important for some of the cases considered below, in which $H$ has a kinematically protected single-particle mode, but $H_0$ does not. In these cases the small perturbation $\lambda V$ leads to the creation of a bound state, which is a non-perturbative effect. In these cases it turns out that the perturbation theory around $H$ as sketched above gives a (qualitatively) correct description of the observed dynamics.

The question we want to address is what happens outside the linear response regime. Linear response theory is usually expected to describe the short-time regime for very small but finite values of $\lambda$, but fail at late times. The question of its regime of applicability is related to the properties of nonlinear response functions, which have recently been analyzed in the class of systems discussed here \cite{fava2022divergent} and shown to acquire late-time divergences in some cases.

The linear response viewpoint summarized above obscures the fact that the quench deposits a finite energy density into the system. A complementary viewpoint on long-lived oscillations is obtained by employing a spectral representation in terms of energy eigenstates of the post-quench Hamiltonian
\begin{equation}
{\rm Tr}\big[{\cal O}\rho(t)\big] =
  \sum_{m,n}e^{i(E_m-E_n)t}\langle m | {\cal O} | n \rangle
\langle n | \rho(t=0) | m\rangle\ .
\label{1pt}
\end{equation}
Assuming that the operator ${\cal O}$ connects states with quasiparticle numbers that differ by one, oscillations with frequency $\Delta_{\rm ex}$ may ensue for the following reason. In
the gas phase energy eigenstates can be viewed as scattering states of
the stable quasiparticles and adding a single quasiparticle with
momentum $q$ to an energy eigenstate (approximately) leads to an energy eigenstate that differs in energy and momentum by $\epsilon(q)$. If $\rho$ is translationally invariant, then the only nonzero contributions to the sum occur at $q=0$. As this process works for all energy eigenstates at
the (low) energy density of interest (which is set by the factor
$\langle n | \rho(t=0) | m\rangle$), one may expect long-lived oscillations in the expectation value \fr{1pt} to occur. 

We stress that the oscillations produced by the mechanism outlined above are not a finite-size effect but can persist in the thermodynamic limit. For all examples discussed below we have verified that varying the system size does not affect the amplitude or frequency of the oscillations observed. This is very different to the oscillations reported in \cite{Reimann2020Temporal}, which indeed are finite-size effects. 

%%%%%%%%%%%%%%%%%%%%%%%%%%%%%%%%%%%%%%%%%%%%%%%%%%%%
\subsection{Haldane-gap chains}
\label{subsec:spin1}
%%%%%%%%%%%%%%%%%%%%%%%%%%%%%%%%%%%%%%%%%%%%%%%%%%%%
As a first example of a model that exhibits undamped oscillations after quantum quenches in the linear response regime we consider the antiferromagnetic bilinear-biquadratic (BLBQ) chain. This is a family of spin-$1$ chains described by \cite{Xian1993Spontaneous,Lauchli2006Spin,Binder2018Infinite} 
\begin{eqnarray}
    H(\gamma) = J\sum_i^{L-1}  \left[\left( \vec{S}_i \cdot
    \vec{S}_{i+1} \right)+ \gamma \left(\vec{S}_i \cdot
    \vec{S}_{i+1}\right)^2 \right] \ . \label{Eq:H_BLBQ} 
\end{eqnarray}
For $\gamma=0$ the model reduces to the spin-1 Heisenberg
antiferromagnet whilst for $\gamma=1/3$ it is the AKLT chain
\cite{Affleck1987rigorous}, whose ground state is an exact MPS with
bond dimension $\chi=2$. 

\begin{figure}[htb]
    \centering
    \includegraphics[scale=0.5]{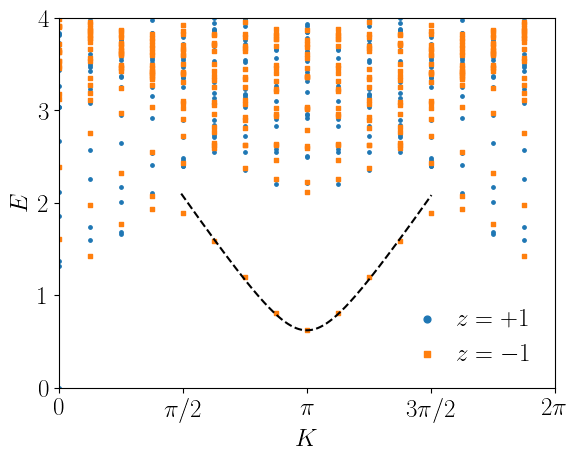}
    \caption{Spectrum of the BLBQ Hamiltonian~\fr{Eq:H_BLBQ} found
      with QuSpin \cite{QuSpinPartI17,QuSpinPartII19} for $L=16$ sites
      and $\gamma=0.25$, within the magnetization
      sector with $S^z_{\rm Tot}=0$. States are coloured by their charge under a
      $\Z_2$ spin-flip symmetry $S^z\mapsto -S^z$, $|\psi\rangle
      \mapsto z |\psi\rangle$. Dashed line is a fit around $k=\pi$ to a functional form $\epsilon(k)=\sqrt{\Delta_{\rm ex}^2+v^2(k-\pi)^2}$.} 
    \label{fig:BLBQ_Spectrum}
\end{figure}

Both values of $\gamma$ lie within the gapped `Haldane
gap phase' \cite{Haldane1983Nonlinear} $-1 < \gamma < 1$. At $\gamma=1$ the model is the $SU(3)$ symmetric Lai-Sutherland model which is gapless \cite{Lai1974Lattice,Sutherland1975Model}. In
Fig.~\ref{fig:BLBQ_Spectrum} we plot the low-energy spectrum obtained
by exact diagonalization (ED) on $L=16$ sites and periodic boundary conditions for $\gamma=0.25$ which
is representative of the Haldane phase, with a gap of $\Delta_{\rm ex}(\pi)\approx 0.62J$ and group velocity $v\approx 1.26J$. 

The ground state is at $k=0$ and there is a gap to a triplet band of
magnons with energy minimum at $k=\pi$.

Consider first global quenches where a finite energy density is
generated by quenching the ratio of exchange constants
$\gamma\rightarrow\gamma'$. For the Hamiltonian~\fr{Eq:H_BLBQ} the
ground state has momentum $k=0$ but the Haldane gap is at $k=\pi$, with the magnon mode only persisting in a region of the Brillouin zone around this that does not extend to $k=0$.
For such a quench both the Hamiltonian and the initial state are translationally invariant, so local operators cannot ``access'' single magnon excitations in the way described above because $\rho_{nm}=\langle n | \rho(t=0) | m \rangle =0$ for energy eigenstates that differ by a single magnon, as they differ in momentum. 
As a result there are no long-lived oscillations for translationally invariant initial states. 
On the other hand these considerations suggest a way out: we
need to choose an initial state that is invariant only under
translations by two sites, which can be achieved simply by choosing the pre-quench Hamiltonian to have an additional staggered magnetic field
\begin{equation}
    H_{\rm pre}= H(\gamma_i) + h_s \sum_m (-1)^m S^z_m  \ . 
\end{equation}
$H_{\rm pre}$ now has a ground state that is invariant only under translation by two sites and $\rho_{nm}\neq 0$ and so as discussed in the previous section we therefore expect to see oscillations at $\epsilon(\pi)$ (the magnon does not extend to $k=0$ so this will be the only frequency present). As a numerical test of these ideas in the BLBQ chain we perform a quench using the ITensor \cite{ITensor} library, which enables us to use DMRG to compute an  approximation to the ground state of $H(\gamma_i,h_s)$, and then to time-evolve the state according to $H(\gamma_f,0)$ using time-evolving block decimation (TEBD). Here and elsewhere, the time window in which we plot TEBD data is determined such that the TEBD results do not change when suitably increasing the bond dimension, in Figs. 2,3 we have plotted data for $\chi=400$ and ensured no change in the corresponding plots for $\chi=600$.

\begin{figure}
    \centering
    \includegraphics[scale=0.5]{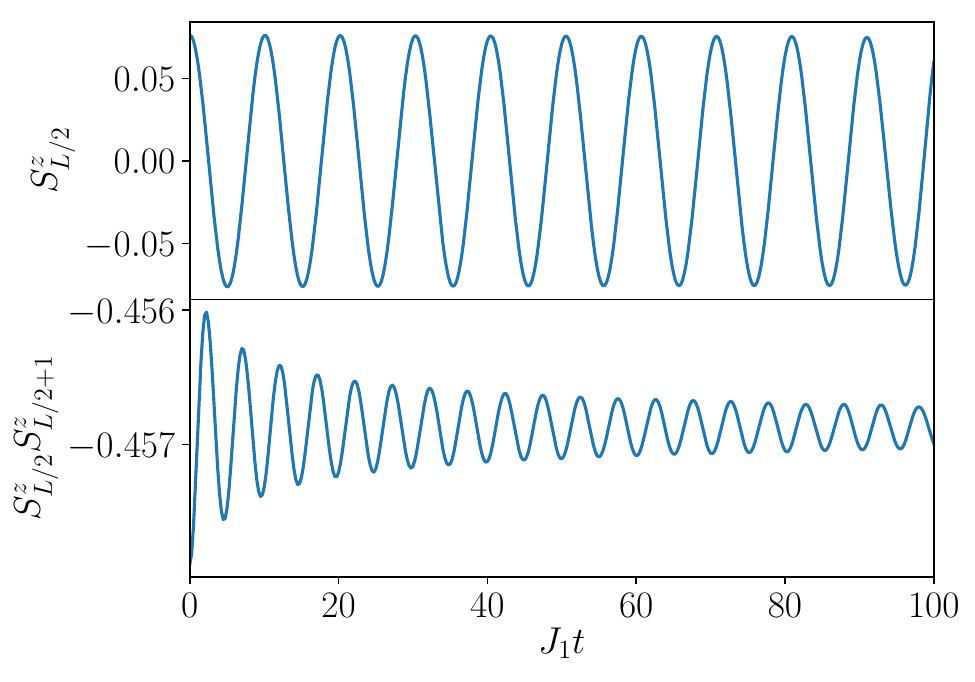}
    \caption{Result of a quantum quench keeping $\gamma=0.25$ and quenching the initial staggered field $h_s:0.01\mapsto 0$. This quench produces an energy per site $\epsilon_{\rm Quench}\approx 3.8 \times 10^{-4}J$ which corresponds to an equilibrium temperature of $T\approx0.12J$ (estimated using ED on 12 sites). TEBD performed using $L=400$ and $\chi=400$. The final time is window is determined by requiring that the results do not change on increasing the bond dimension to $\chi=600$.}
    \label{fig:Spin1-SmallQuench}
\end{figure}

\begin{figure}
    \centering
    \includegraphics[scale=0.5]{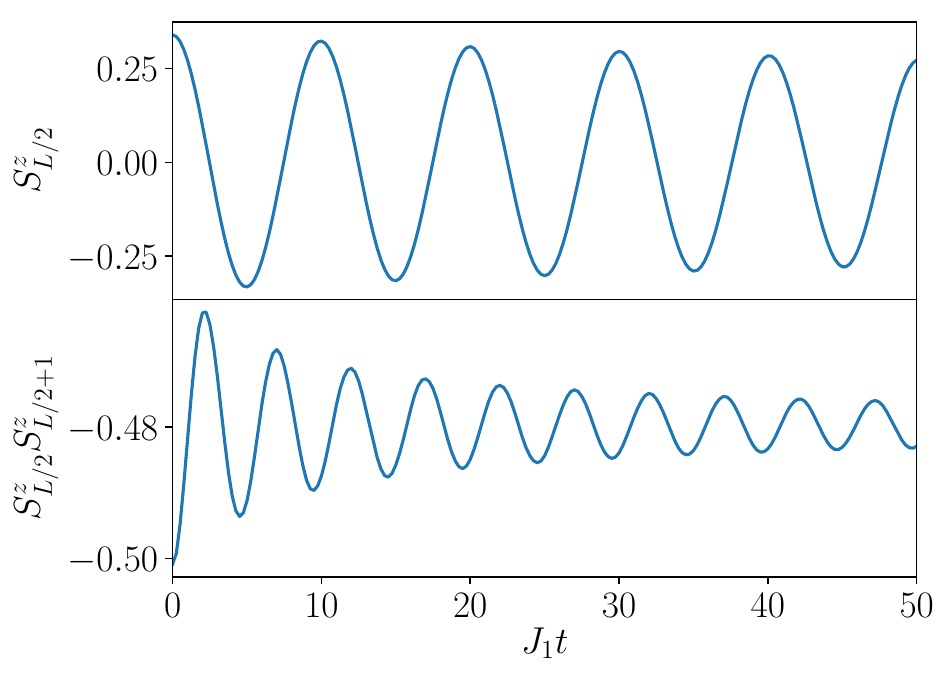}
    \caption{Same as Fig.~\ref{fig:Spin1-SmallQuench} but with an initial staggered field $h_s=0.05$. The energy per site $\epsilon_{\rm Quench}\approx 7.86\times 10^{-3}J$ produced by the quench corresponds to an equilibrium temperature of $T \approx 0.23 J$. TEBD parameters are $L=400, \chi=400$.}
    \label{fig:Spin1-MediumQuench}
\end{figure}

The results are shown in Figs.~\ref{fig:Spin1-SmallQuench},~\ref{fig:Spin1-MediumQuench}
for two quenches of different strengths. For the weaker quench we see that there are oscillations in the quantity $\langle S^z_{L/2}\rangle $ with little to no visible damping, whilst the evolution of $\langle S^z_{L/2}S^z_{L/2+1}\rangle $ is strongly damped. In light of the previous section, this is to be understood as due to a discrete spin-flip symmetry as follows: the oscillations can only occur when the matrix element $\langle {\rm GS} | \mathcal{O} | {\rm QP}(k)\rangle\neq 0$ where $|{\rm QP}(k)\rangle$ is the quasiparticle at momentum $k$ and $|{\rm GS}\rangle$ is the ground state of the post-quench Hamiltonian. For the BLBQ chain the ground state is invariant under $S^z\mapsto -S^z$ but the magnon mode is odd under this symmetry. Therefore, the matrix elements are only non-zero when the observable is $\Z_2$ odd, such as $S^z$ itself, and conversely the oscillations decay rapidly when the observable is $\Z_2$ even. The oscillations in the magnetization have frequency $\omega$ very close to the magnon gap at $k=\pi$ as expected
\be
\omega\approx\ 0.62102 \dots J \ ,\qquad
\Delta_{\rm ex}(\pi)\approx 0.62096 J \ .
\ee

Fig.~\ref{fig:Spin1-MediumQuench} shows a somewhat larger quench from $h_s=0.05$, which still has an energy per site well below the gap. The average inter-particle distance after this quench is approximately
\be
\ell=\Delta_{\rm ex}/\epsilon_{\rm Quench}\approx 79\ .
\ee
From the perspective of a low density gas of quasiparticles, the finite lifetime of the oscillations is caused by scattering events \cite{Birnkammer_2022}. Therefore one would not expect to see appreciable decay at times $vt\lesssim \ell$. From ED we estimate that the group velocity is roughly $v\approx 1.26J$, and so the slight decay by $Jt=50$ makes sense within the quasiparticle gas picture. 
Conversely, for the quench in Fig.~\ref{fig:Spin1-SmallQuench} the mean free path is $\ell \approx 1600$ and thus the lack of decay is also consistent with this rough estimate.
For even larger quenches than in Fig.~\ref{fig:Spin1-MediumQuench} we find that, as expected on the basis of our quasiparticle gas picture, the decay becomes more easily visible at short times.

The quenches in Haldane-gapped models explored above yield several insights - the first is that for weak quenches oscillatory behaviour results as predicted~\cite{Delfino2014Quantum,Delfino2017Theory,Delfino2020Persistent,Delfino2022persistent} when there is a quasiparticle mode. This confirms that such oscillations are unrelated to the formation of bound states or of confinement, except inasmuch as they provide a mechanism for kinematically protected quasiparticles to exist. It also highlights the importance of symmetries which can cause the relevant matrix elements to be zero and relaxation to be consequently much faster. Perhaps most importantly we have evidence that the oscillations do decay, which was suggested in previous studies~\cite{Collura2018Dynamical,Robinson2019Signatures} but not observed in the models considered therein. Our findings also show that in the case studied above the regime of validity of the perturbative approach used in ~\cite{Delfino2014Quantum,Delfino2017Theory,Delfino2020Persistent,Delfino2022persistent} is limited to short times.

%%%%%%%%%%%%%%%%%%%%%%%%%%%%%%%%%%%%%%%%%%%%%%%%%%%%
\subsection{\label{subsec:XXZ}Dimerized XXZ Model}
%%%%%%%%%%%%%%%%%%%%%%%%%%%%%%%%%%%%%%%%%%%%%%%%%%%%
Generally the quasiparticle gas can consist of several species, for instance in models with ``elementary'' quasiparticle excitations as
well as (multi-particle) bound states. 
We have already mentioned two examples of this situation - the Ising model with both longitudinal and transverse field
\cite{KormosRealtime2016,Delfino2017Theory,Collura2018Dynamical,Robinson2019Signatures} and the Ising model with transverse field only, but additional next-nearest neighbour Ising interactions \cite{Robertson2023Simple}.
Neither of these lattice models analyzed in the context of persistent
oscillations exhibits a $\mathsf{U}(1)$ symmetry associated with particle
number conservation, which greatly complicates the application of
perturbative approaches based on the BBGKY hierarchy or the flow
equation approach
\cite{moeckel2009real,moeckel2010crossover,essler2014quench,Hermanns2013Few,Bertini15Prethermalization,Bertini16Thermalization,nessi2014equations,Nessi2015glass}.
In order to study the fate of these oscillations at very late times we
therefore introduce a spin-1/2 dimerized XXZ model in a staggered
field, which can be mapped to a model of spinless interacting fermions with
particle number conservation. This enables us to apply the equations
of motion techniques developed in
\cite{Bertini15Prethermalization,Bertini16Thermalization}. This model
features elementary fermionic excitations as well as bosonic
two-particle bound states. Moreover, in the appropriate scaling limit the model reduces
to the sine-Gordon quantum field theory in the attractive regime. The Hamiltonian of our model is 
\begin{align}
H(\Delta,\alpha,&h_s) 
= -\frac{J}{2} \sum_{m=0}^{L-1} [1+\alpha(-1)^m][S^+_m S^-_{m+1}+{\rm h.c.}]  \nn
& + \Delta J\sum_m S^z_m S^z_{m+1} + h_s J\sum_m (-1)^m S^z_m \ . 
 \label{Eq:HSpin}
\end{align}
Here $\alpha$ tunes the degree of dimerization in the XY plane
and $h_s$ is a staggered applied field. For $\alpha=h_s=0$ and $|\Delta|<1$ the model
reduces to the integrable spin-1/2 XXZ chain in the massless Luttinger
liquid phase, whilst non-zero values of $\alpha,h_s$ break the
integrability and open a gap in the zero magnetization sector. A Jordan-Wigner transformation maps the
Hamiltonian~\fr{Eq:HSpin} to one of interacting spinless fermions with
interaction strength $\Delta$. As in the case of the BLBQ chain, discrete symmetries play a role in allowing or disallowing persistent oscillations. When only one of $\alpha, h_s$ is present in (\ref{Eq:HSpin}) there is a discrete spatial $\Z_2$ symmetry corresponding to reflection across a bond or site, respectively. We will elaborate the importance of retaining both parameters after presenting data from quenches.

\begin{figure}[t]
\centering
\includegraphics[scale=0.5]{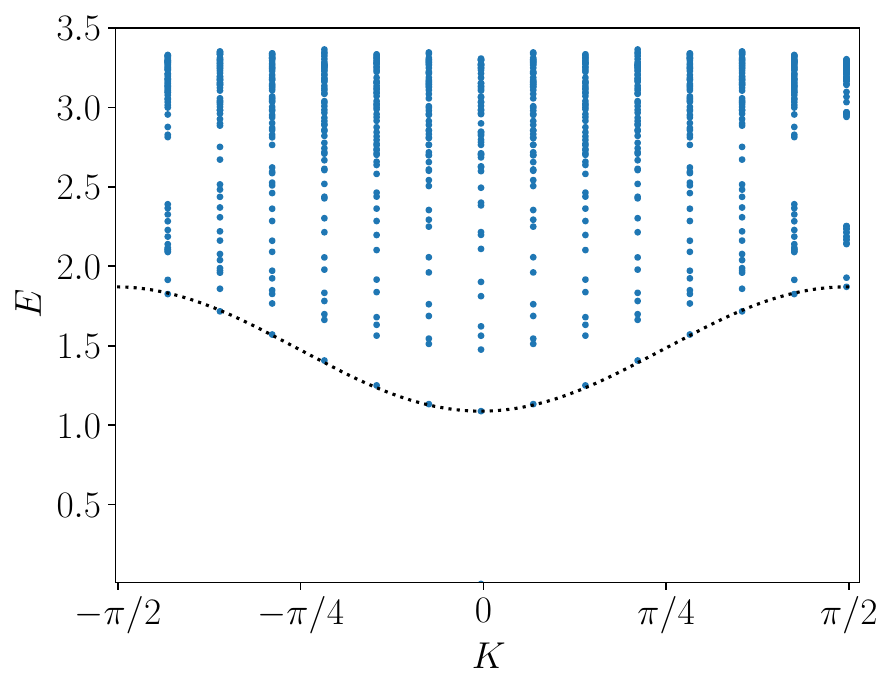}
\caption{Low energy spectrum of the Hamiltonian \fr{Eq:HSpin} in the $S^z_{\rm Tot}=0$ sector for $\alpha=0.4, h=0.2, \Delta=0.65$, calculated for $L=28$ spins. The dotted black curve is a fit of the bound state to $\epsilon(k)=M-\frac{v}{2}(\cos 2k -1)$ where $v\approx 0.78$ is the maximal group velocity.}
\label{fig:EDSpectrumXXZ}
\end{figure}

Rotational symmetry about the $z$ axis corresponds to $\mathsf{U}(1)$ particle number conservation in the fermionic variables. 
In the low-energy limit, the Hamiltonian (\ref{Eq:HSpin}) for $|\Delta|<1$ reduces to a sine-Gordon model \cite{Essler2005Application}, whose low lying excitations for $\Delta>0$ are solitons, anti-solitons and soliton-antisoliton bound states known as `breathers'. 
In the lattice model we determine the spectrum of low-lying excitations in the $S^z_{\rm Tot}=0$ sector  by exact diagonalization, \emph{cf.} Fig.~\ref{fig:EDSpectrumXXZ}.
We can see that throughout the Brillouin zone there is a bound state visible below a continuum of states. The bound state is kinematically protected and therefore stable. This establishes that our model fulfils the first of our requirements.

We investigate the time evolution using both TEBD, which is capable of exactly describing the evolution of states with sufficiently low entanglement, and perturbative methods which are valid at small $\Delta$. The latter is needed because following a quench the entanglement entropy generically grows linearly in time \cite{Kim2013ballistic,Nahum2017Quantum,Zhou2020Entanglement}. As such the true time evolved state quickly leaves the manifold of states that can be accurately described by matrix-product states with finite bond dimensions. We adopt open boundary conditions when performing TEBD numerics; when working with the equivalent fermionic model it suffices to work in the sector with fixed fermion number and we adopt periodic boundary conditions out of convenience. We also ensure that our system sizes are sufficiently large to rule out
finite-size effects such as traversals \cite{EsslerQuench2016} on the time scales we are interested in. 

To ensure a long window of applicability of the perturbative approaches \cite{Bertini15Prethermalization,Bertini16Thermalization}
we consider quantum quenches from an initial thermal state of the
non-interacting model $\rho(0,\alpha,h_s,\beta)$, where 
\begin{equation}
    \rho(\Delta, \alpha,h_s,\beta) = \frac{\exp\left(-\beta H(\Delta,\alpha,h_s)\right)}{\Tr \left[\exp\left(-\beta H(\Delta,\alpha,h_s)\right)\right]} \ .
\end{equation}
As the perturbative approaches expand around thermal states of free Hamiltonians, they are able to describe states with volume law entanglement, unlike MPS methods. Instead, they are limited by their assumption that higher particle cumulants are negligible.
This is then time evolved using the Schr\"{o}dinger equation for
$H(\Delta,\alpha',h'_s)$. We focus on expectation values of
local observables such as the staggered magnetization within a unit cell and nearest-neighbour spin-bilinears
\begin{align}
    m_{s} &= \langle \psi(t) | S^z_{2n} - S^z_{2n+1} | \psi(t) \rangle ,\nn
    S^{\alpha \alpha}_{m,m+1}&=\langle S^\alpha_m S^\alpha_{m+1} \rangle \ .
\end{align}
\begin{figure}[h]
    \centering
    \includegraphics[scale=0.50]{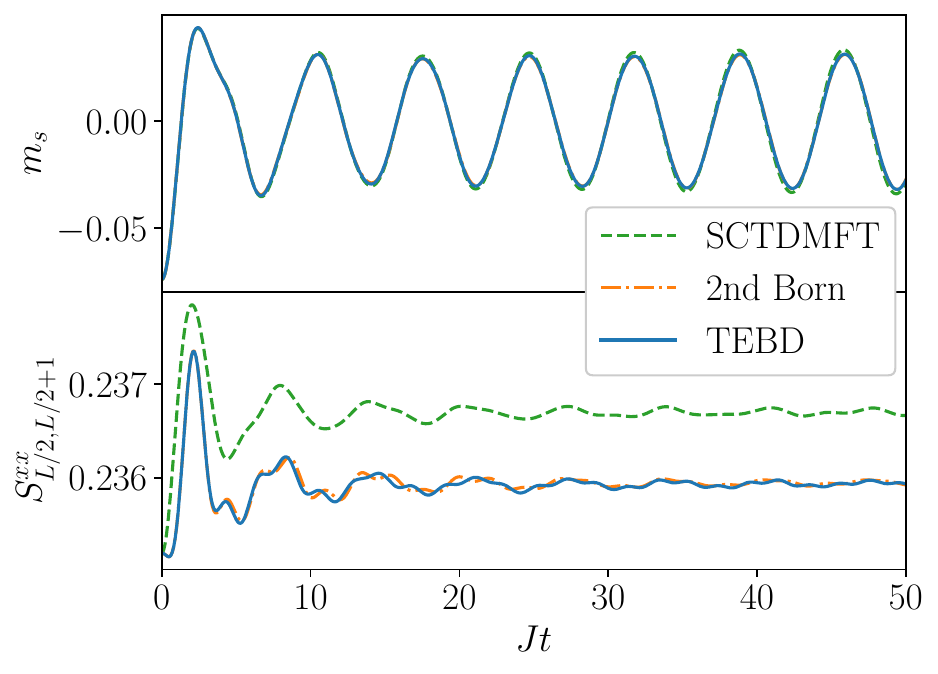}
    \caption{Time evolution following a quench from the ground state with $\Delta:0\mapsto 0.2, h_s:0.1\mapsto 0$ and $\alpha_s=0.4$ before and after the quench. Upper: staggered magnetization, showing persistent oscillations. Lower: Time evolution of $\langle S^x_m S^x_{m+1}\rangle$ where $m=L/2$. Calculations use $L=400$ sites and a maximum bond dimension of $\chi=1000$ for the TEBD.} 
    \label{fig:TEBD_NN}
\end{figure}   

We plot these quantities in Fig.~\ref{fig:TEBD_NN} computed using TEBD on $L=400$ sites with the ITensor \cite{ITensor} library, and two approximate calculations, self-consistent time-dependent mean field theory and the second Born Approximation, both detailed in Section \ref{Sec:Decay}. Fig.~\ref{fig:TEBD_NN} shows the case where the time evolution has a $\Z_2$ symmetry (reflection in a bond as $h'_s=0$) but the initial state does not. 
In this case, the staggered magnetization initially has a non-zero value and then oscillates about the thermal value of $0$. The frequency agrees with the energy difference between the post-quench ground state and the bound state, which have opposite $\Z_2$ parities and are thus connected by the $\Z_2$ odd operator $m^z_s$.
Conversely, the operator $S^{xx}_{m,m+1}$ is $\Z_2$ even and thus has decaying expectation value at late times.
We note that in the bottom panel of Fig.~\ref{fig:TEBD_NN} the SCTDMFT does not appear to accurately capture the early time evolution. This is despite the fact that SCTDMFT is expected to be accurate at early times, as will be explained in Section~\ref{Sec:MFT}. We have checked that the difference between it and the second Born approximation scales like $\mathcal{O}(\Delta^2)$ and so this deviation indicates that for the observable in question the $\mathcal{O}(\Delta)$ contribution is either very small or absent.

%%%%%%%%%%%%%%%%%%%%%%%%%%%%%%%%%%%%%%%%%%%%%%%%%%%%%%
\section{\label{Sec:Decay}Decay of oscillations in the staggered XXZ  model at late times}
%%%%%%%%%%%%%%%%%%%%%%%%%%%%%%%%%%%%%%%%%%%%%%%%%%%%%%
We now turn to the ``intermediate" time regime. This is no longer accessible to TEBD for the reasons set out above, but can be studied by appropriate truncation schemes of the BBGKY hierarchy. We first consider the simplest such scheme, a self-consistent time-dependent mean field theory \cite{boyanovsky1998evolution,sotiriadis2010quantum,Collura20Order,vannieuwkerk2019self,*vannieuwkerk2020on,vannieuwkerk2021Josephson,lerose2019impact,Robertson2023Simple}.
%%%%%%%%%%%%%%%%%%%%%%%%%%%%%%%%%%%%%%%%%%%%%%%%%%%%%%
\subsection{\label{Sec:MFT}Self-consistent time-dependent mean-field theory (SCTDMFT)}
%%%%%%%%%%%%%%%%%%%%%%%%%%%%%%%%%%%%%%%%%%%%%%%%%%%%%%
The correct degrees of freedom for constructing a mean-field approximation for \fr{Eq:HSpin} are fermions rather than spins. Applying a Jordan-Wigner transformation to the Hamiltonian gives
\begin{eqnarray}
    H(\Delta,\alpha,h_s) = -\frac{J}{2}\sum_{m=0}^{L-1}(1+\alpha(-1)^m)(c^\dag_m c_{m+1}+\text{h.c}) \nn 
    +\Delta\sum_m n_mn_{m+1} - h_s\sum_m (-1)^m n_m \ ,\ \label{Eq:HFerm}
\end{eqnarray}
where $c_n$ are spinless fermions and $n_m=c^\dagger_mc_m$. If the fermion parity is odd the $c_n$ have periodic (Ramond) boundary conditions whilst if the fermion parity is even they have anti-periodic (Neveu-Schwarz) boundary conditions. We work at half filling such that these conditions correspond to the number of unit cells being odd/even respectively. In SCTDMFT the interaction terms in Eq.~\fr{Eq:HFerm} are decoupled in a time-dependent way, which corresponds to normal-ordering with respect to the time-evolving state of the system and retaining only terms that are quadratic in fermionic creation/annihilation operators
\begin{align}
n_mn_{m+1} \mapsto &-\langle c_m^\dag c_{m+1} \rangle_t  c_{m+1}^\dag c_m - c_m^\dag c_{m+1} \langle c_{m+1}^\dag c_m \rangle_t\nn
&+ \langle n_m \rangle_t n_{m+1}
+ n_m\langle n_{m+1} \rangle_t \nn 
&+ |\langle c_m^\dag c_{m+1}\rangle_t|^2 - \langle n_m
\rangle_t \langle n_{m+1}\rangle_t \ .  
\end{align}
Here the expectation values $\langle .\rangle_t$ in the time-evolving state are determined self-consistently. This produces a mean-field
Hamiltonian $H_{\rm MFT}(t)$ with the following key properties: 
\begin{enumerate}
\item $H_{\rm MFT}(t)$ is quadratic at all times and thus time evolving a Gaussian state with $H_{\rm MFT}(t)$ ensures that Wick's theorem holds at all times.
\item The equations of motion for the two-point functions obtained
      using $H_{\rm MFT}$ agree with the equations of motion obtained
      using the full Hamiltonian, under the approximation that Wick's theorem is valid.
\end{enumerate}
Since our initial state is a thermal state of the free fermion
Hamiltonian $H(0,\alpha,h_s)$ it is Gaussian and the SCTDMFT is thus accurate at early times by construction. The time-dependent mean-field Hamiltonian takes the form
\begin{align}
    H_{\rm MFT}(t) =& -\sum_{s=0}^{\frac{L}{2}-1}\Big[\frac{J_0(t)}{2} c_{s,0}^\dag c_{s,1} + 
    \frac{J_1(t)}{2} c_{s,1}^\dag c_{s+1,0}+{\rm h.c.}\nn
    & + h_{\rm eff}(t)\big( c^\dag_{s,0}c_{s,0}-c^\dag_{s,1}c_{s,1}\big)\Big] +E_0(t) \ , 
    \label{Eq:HMFT}
\end{align}
where $s$ now labels the unit cell and $a=0,1$ the sites within it
such that the spin labelled $(s,a)$ is at position $m=2s+a$ in the
chain. The constant term $E_0(t)$ does not have any effect on the
equations of motion, but ensures that the expectation value of energy
is conserved. The mean-field Hamiltonian contains the effective
couplings  
\begin{eqnarray}
J_0(t) &=& J(1+\alpha)+\Delta \langle c^\dag_{s,1}c_{s,0}\rangle_t \ , \nn 
J_1(t) &=& J(1-\alpha)+\Delta \langle c^\dag_{s+1,0}c_{s,1} \rangle_t \ , \nn 
h_{\rm eff}(t) &=& h_s - \Delta\langle c^\dag_{s,0}c_{s,0}-c^\dag_{s,1}c_{s,1} \rangle_t \ . 
\end{eqnarray}
Note that $J_{0,1}(t)$ are generically complex at intermediate times
and that $h_{\rm eff}$ is (up to a constant shift and rescaling) equal
to the staggered magnetization $m_s(t)$.
The mean-field Hamiltonian can be block-diagonalized by the canonical
transformation 
\begin{eqnarray}
    c_{s,a}=\frac{1}{\sqrt{L}}\sum_s e^{ik(2s+a)}(c_{+}(k)+(-1)^a c_{-}(k)) \ , 
\end{eqnarray}
where $k=2\pi n/L$ for $n=0,\dots L/2-1$. We find
\begin{align}
H_{\rm MFT}(t) &= -\sum_k \sum_{\mu\nu\in\{+,-\}}\Tilde{h}^{\rm MFT}_{\mu\nu}(k,t)c_\mu^\dag(k) c_\nu(k) \ , \nn
    \Tilde{h}^{\rm MFT}_{\mu\nu}(t) &= \begin{pmatrix} A(k,t) & B(k,t) - i B'(k,t) 
    \\ B(k,t) + i B'(k,t) & -A(k,t)\end{pmatrix} \ . \label{Eq:HMFT_BlockDiag}
\end{align}
Here $A(k,t), B(k,t)$ and $B'(k,t)$ are real functions that can be
expressed in terms of $J_\pm(t)=J_0(t)\pm J_1(t)$ and $h_{\rm eff}(t)$
as
\begin{eqnarray}
    A(k,t) & = & \Re\big(J_+(t)\big) \cos k - \Im \big(J_+(t)\big) \sin k \ , \nn 
    B(k,t) & = & h_{\rm eff}(t) \ , \nn 
    B'(k,t) & = & \Re\big(J_-(t)\big) \sin k + \Im \big(J_-(t)\big) \cos k \ . 
\end{eqnarray}

The time evolution of the momentum space two-point functions $n_{\mu
  \nu}(k)=\langle c_\mu^\dag(k) c_\nu (k) \rangle, \ \mu,\nu\in \{+,-\}$ is
now easily obtained from the Heisenberg equations of motion
\begin{eqnarray}
     \frac{{\rm d}n_{++}(k,t)}{{\rm d}t} &=& 2B'\Re(n_{+-}) - 2B \Im(n_{+-})\ , \nn
     \frac{{\rm d}n_{--}(k,t)}{{\rm d}t} &=& -2B'\Re (n_{+-}) + 2B \Im(n_{+-}) \ , \\ 
     \frac{{\rm d}n_{+-}(k,t)}{{\rm d}t} &=& (B'-iB)(n_{--}-n_{++})-2iA(k)n_{+-} \ . \nonumber 
\end{eqnarray}
These equations can alternatively be derived by a first order truncation of the BBGKY hierarchy, \emph{cf.} Ref.~\cite{Bertini16Thermalization}
\begin{figure}[!h]
    \centering
    \includegraphics[scale=0.5]{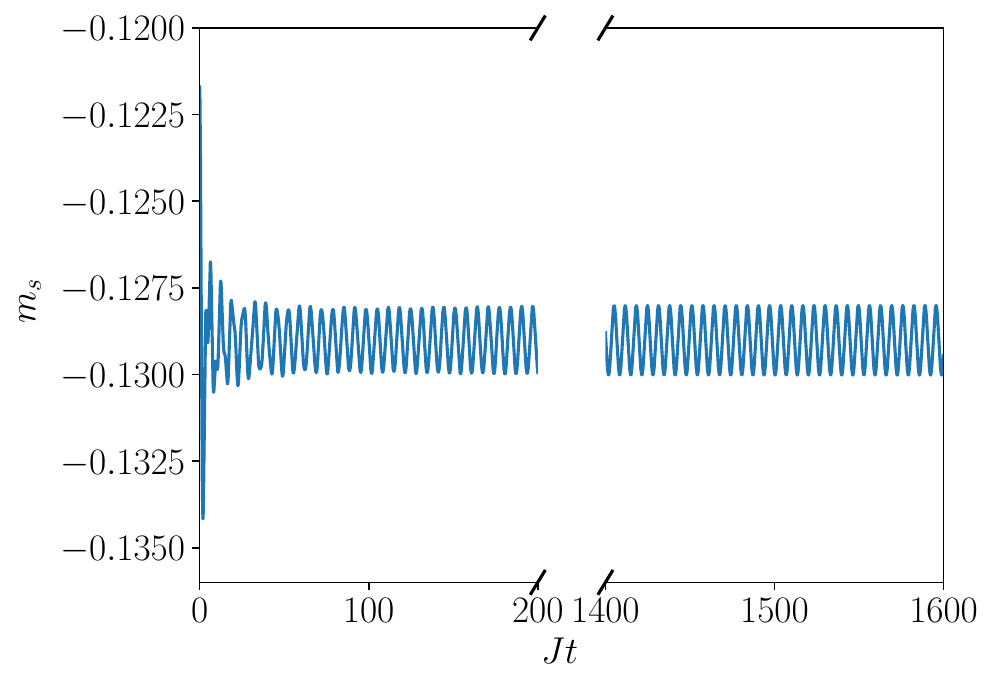}
    \caption{Mean-field evolution of the staggered magnetization after a quench with initial state
      $\rho(0,0.4,0.2,4.0)$ and time evolved using the Hamiltonian $H(0.1,0.4,0.2)$. The
      oscillations are undamped at late times in this approximation up
      to the light-cone time set by the system size (here
      $L=2000$).}
    \label{fig:MFT}
\end{figure}

We solve these equations of motion numerically, updating the mean fields $J_\pm(t)$ and $h_{\rm eff}(t)$ every timestep and plot the resulting staggered magnetization following a quench from an initial thermal state in Fig.~\ref{fig:MFT}, which shows clear oscillations that become highly monochromatic and undamped at late times. The choice of thermal state is made to ensure that the system has an energy per site well above the ground state but still small compared to the gap. This ensures that there is a low density of quasiparticles in the system and that they can be treated as a dilute gas. 

\begin{figure}[h]
    \centering
    \includegraphics[scale=0.5]{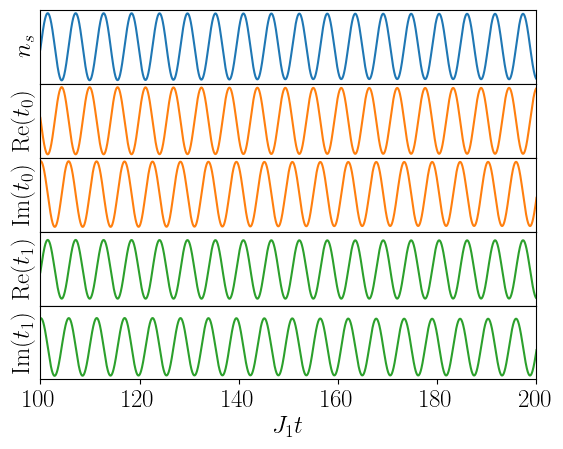}
    \caption{Evolution of the mean fields $t_0=\langle c^\dag_{s,0}c_{s,1}\rangle, t_1=\langle c^\dag_{s,1}c_{s+1,0}\rangle, n_s=\frac{1}{2}\langle c^\dag_{s,0}c_{s,0}-c^\dag_{s,1}c_{s,1}\rangle$ following a quench from the initial state $\rho(0,0.4,0.3,\infty)$ and evolved with the Hamiltonian $H(0.3,0.4,0.3)$.}
    \label{fig:XXZMeanFieldsFFT}
\end{figure}

We now return to the physical origin of the oscillations and their frequency. To that end we have considered ground state quenches at $\alpha=0.4, h_s=0.3$, i.e. initial density matrices $\rho(0,0.4,0.3,\infty)$, for several $\Delta$. In this case we observe essentially a single oscillation frequency $\omega_B$ at intermediate and late times. For example, in Fig.~\ref{fig:XXZMeanFieldsFFT} we show the evolution of the mean fields.
\begin{figure}[hb]
    \centering
    \includegraphics[scale=0.5]{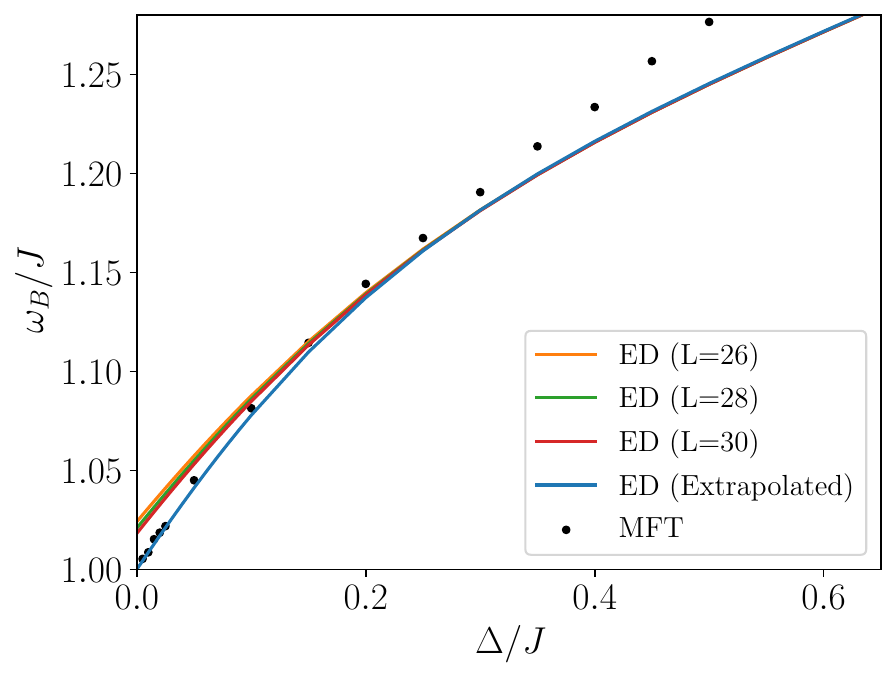}
    \caption{Estimation of the bound state mass using ED compared to the persistent frequency extracted from the mean-field evolution of the initial state
      $\rho(0,0.4,0.3,\infty)$ by the Hamiltonian $H(\Delta,0.4,0.3)$. The ED exhibits large finite size effects when $\Delta\to 0$ so we plot the result of extrapolating to $L=\infty$ by fitting the gap to a power series in $1/L^2$ up to $L^{-4}$. }
    \label{fig:EDvsMFT_BSMss}
\end{figure}
Performing a fast Fourier transform using data from $t=50$ up to $t=2000$ gives a single sharp peak at the frequency $\omega_B$. We compare this to the energy of the first excited state computed by exact diagonalization on system sizes up to $L=30$ in Fig.~\ref{fig:EDvsMFT_BSMss}. We observe that the oscillation frequency observed in SCTDMFT is in very good agreement with the bound state gap at $q=0$ for small interaction strengths $\Delta\alt 0.25J$. For small interactions strengths $\Delta\approx 0$ the ED results exhibit sizeable finite size effects that are discussed in Appendix \ref{App:Delta0}. Using the extrapolation procedure summarized there provides us with the curve labelled `ED (Extrapolated)' in Fig.~\ref{fig:EDvsMFT_BSMss}.

The emergence of persistent oscillations of the expectation values of observables in the SCTDMFT can be understood as follows. The solutions to the self-consistency equations reported above exhibit periodic behaviour with a single frequency. As a result the SCTDMFT is equivalent to a periodically driven system with a Hamiltonian that is quadratic in fermions. It is well known that such systems typically synchronize at late times and physical observables then display persistent oscillations at the driving frequency \cite{lazarides2014equilibrium}.

\subsection{\label{Sec:2BA}Second Born approximation}

The lack of damping in SCTDMFT is in fact not surprising as the method
is perturbative to first order in $\Delta$ (at the level of the
equations of motion). In thermal equilibrium we have to evaluate the
self-energy to second order in $\Delta$ in order to obtain a
non-vanishing imaginary part that signals a finite life-time of the
fermions. This suggests that finite lifetime effects in the
non-equilibrium setting of interest here can be captured by the
`second Born approximation'
\cite{Hermanns2013Few,Bertini15Prethermalization,Bertini16Thermalization}. 
We follow \cite{Bertini15Prethermalization} in deriving the equations of
motion for fermionic bilinears $\hat{n}_{\mu\nu}(k,t)=\hat{b}^\dag_\mu(k,t) \hat{b}_\nu(k,t)$, where $\hat{b}^\dag_\mu(k)$ are Bogoliubov fermions 
\begin{equation}
    \hat{c}_m = \frac{1}{\sqrt{L}}\sum_{k>0}\sum_{\mu=0,1}\gamma_\mu(m,k ; \alpha,h_s) \hat{b}_\mu(k) \ ,
\end{equation}
chosen to diagonalize the quadratic part of the Hamiltonian
\begin{align}
\hat{H}(\Delta,\alpha,h_s)&=\sum_{k>0,\mu}\epsilon_\mu(k)\hat{b}_\mu^\dag(k) \hat{b}_\mu(k) \ , \nn 
&+\Delta \sum_{\vec{\mu}}\sum_{k_1,\dots k_4>0} V_{\vec{\mu}}(\vec{k})\hat{A}_{\vec{\mu}}(\vec{k}) \ , 
\end{align}
where we have introduced shorthand notations $\vec{k}=(k_1,k_2,k_3,k_4), \vec{\mu}=(\mu_1,\mu_2,\mu_3,\mu_4)$ and $\hat{A}_{\vec{\mu}}(\vec{k})=b^\dag_{\mu_1}(k_1)b^\dag_{\mu_2}(k_2)b_{\mu_3}(k_3)b_{\mu_4}(k_4)$. Explicit expressions for $\gamma_\mu(m,k|\alpha,h_s)$ and $V_{\vec{\mu}}(\vec{k})$ are given in Appendix~\ref{App:2BA}. 

The equations of motion to second order in $\Delta$ for $n_{\mu\nu}=\langle\hat{n}_{\mu\nu}(k,t)\rangle$ are obtained by truncating the BBGKY hierarchy as derived in \cite{Bertini15Prethermalization, Bertini16Thermalization}. The result is
\begin{widetext}
    \begin{eqnarray}
        \partial_t n_{\mu\nu}(k)&=&i\epsilon_{\mu_\nu}(k) n_{\mu\nu}(k) + 4i\Delta\sum V_{\mu_1\mu_2\mu_3\mu}(k,q,q,k) e^{i(\epsilon_{\mu_1\nu}(k)+\epsilon_{\mu_2\mu_3}(q)t}n_{\mu_1\nu}(k,0)n_{\mu_2\mu_3}(q,0) \nn 
        && -4i\Delta \sum V_{\nu \mu_2 \mu_3 \mu_1}(k,q,q,k)e^{i(\epsilon_{\mu\mu_1}(k)+\epsilon_{\mu_2\mu_3}(q))t}n_{\mu\mu_1}(k,0)n_{\mu_2\mu_3}(q,0) \nn 
        && -\Delta^2 \int_0^t {\rm d}t' \sum K_{\mu\nu}^{\vec{\gamma}}(k_1,k_2;k;t-t')  n_{\gamma_1,\gamma_2}(k_1,t')  n_{\gamma_3,\gamma_4}(k_2,t')      \label{Eq:2BA_EoM}\ \nn
        && -\Delta^2 \int_0^t {\rm d}t' \sum L_{\mu\nu}^{\{\alpha_i\}}(k_1,k_2,k_3;k;t-t') n_{\alpha_1,\alpha_2}(k_1,t')  n_{\alpha_3,\alpha_4}(k_2,t') n_{\alpha_5,\alpha_6}(k_3,t')  \ ,
    \end{eqnarray}
     where the kernel functions $L_{\mu\nu}, K_{\mu\nu}$ are given by 
    \begin{eqnarray}
    K^{\bm\gamma}_{\mu\nu}(k_1,k_2;k;t) &=& 4 \sum_{k_3,k_4{>0}} \sum_{\eta,\eta'} 
    X^{\gamma_1\gamma_3\eta\eta';\eta\eta'\gamma_4\gamma_2}_{{\bm k};{\bm k}'} (\mu,\nu;k;t),\nn
   L^{\{\alpha_i\}}_{\mu\nu}(k_1,k_2,k_3;k;t) &=& 
    8 \sum_{\eta}\sum_{k_4{>0}} X^{\alpha_1\alpha_3\alpha_6\eta;\eta\alpha_5\alpha_4\alpha_2}_{{\bm k};{\bm k}'}(\mu,\nu;k;t)
    - 16\sum_{\eta} X^{\alpha_1\alpha_3\eta\alpha_4;\alpha_5\eta\alpha_6\alpha_2}_{k_1k_2k_1k_2;k_3k_1k_3k_1}(\mu,\nu;k;t)~,\nn
    X^{{\bm\gamma};{\bm\eta}}_{{\bm k};{\bm q}}(\mu,\nu;q;t) &=&
    Y^{\bm\gamma}_{\mu\nu}({\bm k},q)V_{\bm\eta}({\bm q}) e^{i
      E_{\bm\gamma}({\bm k})t} - ({\bm \gamma},{\bm k})\leftrightarrow({\bm
      \eta},{\bm q}) \ , \label{Eq:kernelsEOM}
    \end{eqnarray}
    and where $E_{\vec{\gamma}}(\vec{k}) = \epsilon_{\gamma_1}(k_1) + \epsilon_{\gamma_2}(k_2) - \epsilon_{\gamma_3}(k_3) - \epsilon_{\gamma_4}(k_4)$. 
\end{widetext}
The derivation of Eq.~\fr{Eq:2BA_EoM} is summarized in Appendix~\ref{App:2BA}. We note that $n_{\mu\nu}$ are different from the quantities $n_{\pm\pm}(k,t)$ considered in the SCTDMFT of the previous section. Taking this into account one sees that the SCTDMFT agrees with \fr{Eq:2BA_EoM} up to order $\mathcal{O}(\Delta^1)$ and disagrees with the $\mathcal{O}(\Delta^2)$ terms as expected. Solving Eq.~\fr{Eq:2BA_EoM} requires a runtime of $\mathcal{O}(L^3\times T)$ where $T$ is the simulation time reached. Since simulating up to time $T$ requires a system size at least $2v_{\rm LR}T$ where $v_{\rm LR}$ is the Lieb-Robinson velocity, investigating up to time $T$ scales as $\mathcal{O}(T^4)$.

    The second Born approximation is premised on the assumption that many-particle cumulants are small at intermediate times, whilst going beyond SCTDMFT by allowing for a non-Gaussian state. A priori, this is an uncontrolled approximation. However, we start in a Gaussian state in which all cumulants vanish and so the approximation must be accurate for early times, and becomes better as the interaction strength $\Delta$ becomes small. Furthermore, it gives rise to a Boltzmann equation at late times \cite{Bertini15Prethermalization,Bertini16Thermalization} which is believed to become exact in the ``Boltzmann scaling limit''. This suggests that the approximation remains accurate on time scales $t\sim \Delta^{-2}$. At intermediate times one expects the second Born approximation to continue to provide useful physical insights even if it may not retain full quantitative accuracy. What we wish to establish in this paper is that whilst the SCTDMFT treatment agrees with the prediction of \cite{Delfino2014Quantum,Delfino2017Theory,Delfino2020Persistent,Delfino2022persistent}, the leading correction provided by the second Born approximation causes the oscillations to damp. Finally, we note that the second Born approximation is complementary to TEBD in such cases, since the former can reproduce volume law entanglement but cannot fully capture strong interactions, whilst the latter method can only describe states with a finite amount of entanglement related to the bond dimension used but can describe strongly correlated states at sufficiently low entanglement.

\begin{figure}[ht]
    \centering
    \includegraphics[scale=0.50]{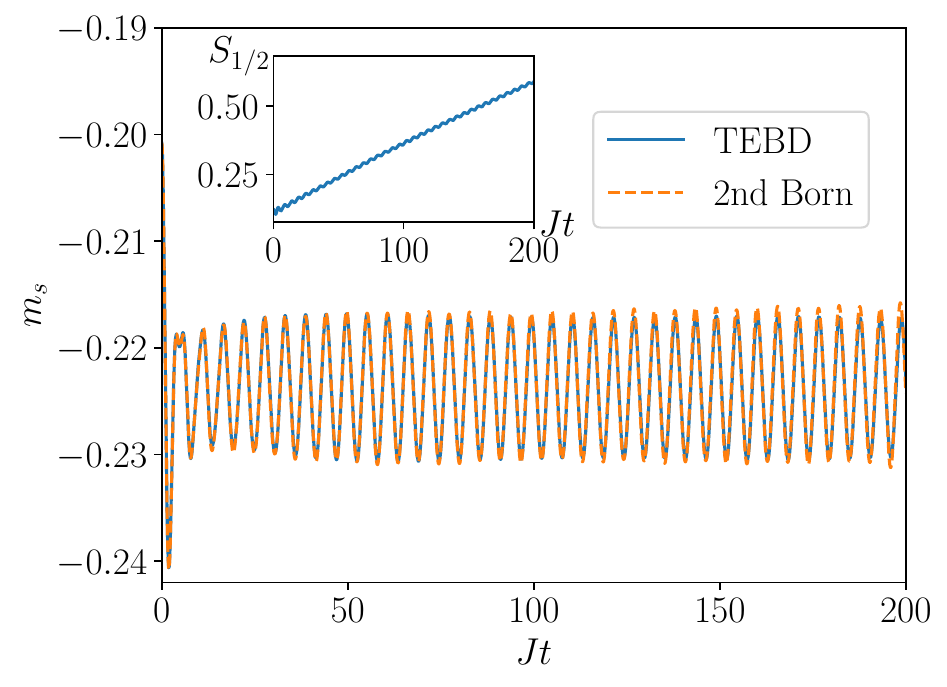}
    \caption{Staggered magnetization for a ground state quench from $\rho(0,0.4,0.3,\infty)$ and time evolving with $H(0.2,0.4,0.3)$. This corresponds to a very small quench with quasiparticle density $\sim 1.6\times 10^{-3}$. Accordingly very large system sizes and late times would be required to observe decay of oscillations. Inset: half chain entanglement entropy, which grows linearly for all times shown. TEBD calculation done with maximum bond dimension $\chi=800$.}
    \label{fig:SBAComp}
\end{figure}

In order to clearly exhibit some of the issues associated with the damping of oscillations we first consider ground state quenches, in which we initialize the system in the ground state of $H(0,0.4,0.3)$ and time-evolve with $H(0.2,0.4,0.3)$ for a system size of $L=300$. Fig.~\ref{fig:SBAComp} shows the time evolution of the staggered magnetization and observe long-lived oscillations with no apparent damping on the long time scales considered. 
This is however entirely expected as the quench produces a very small energy per site  $\epsilon_{\rm quench}\approx 0.0019J$ whilst the gap to create a quasiparticle is $\Delta_{\rm ex}\approx 1.18J$. The resulting average inter-particle distance is therefore $\ell=\Delta_{\rm ex}/\epsilon_{\rm Quench}\approx 640$. That this exceeds the system size simulated means that finite size effects such as traversals will matter long before the many-body effects that would dampen the oscillations. 
The fact that we are effectively dealing with the linear response regime is also apparent from the fact that TEBD is able to access very large time scales $Jt\sim 100$, which means that the volume-law contribution to the entanglement entropy is still negligible.

These considerations show that the energy per site deposited by the quench should be small compared to the bound state energy, however it should not be so small that quasiparticle interactions are negligible on accessible time scales. To overcome this problem we consider larger quenches of the interaction parameter as well as thermal initial states, which provide us with a simple parameter --- the pre-quench temperature --- to vary the post-quench energy density.
For finite pre-quench temperatures we cannot use TEBD since the initial state has volume law entanglement, and so only show the second Born and SCTDMFT results.
We compare the results obtained by the second Born approximation to  SCTDMFT, which as discussed before exhibits persistent oscillations at a frequency that is very close to the bound state energy gap.
\begin{figure}[ht]
    \centering
    \includegraphics[scale=0.5]{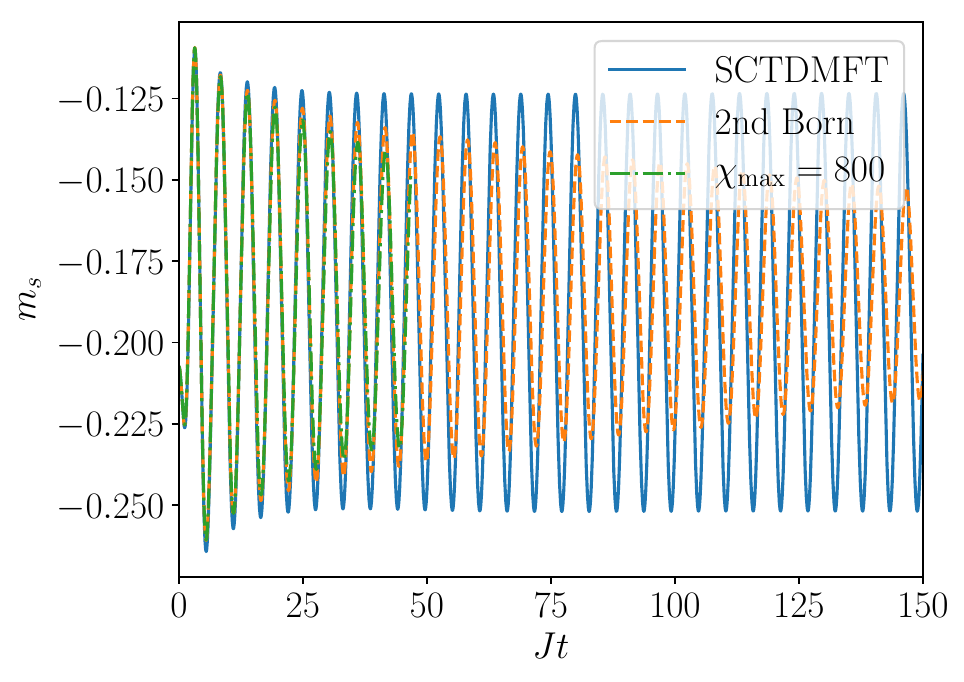}
    \caption{Staggered magnetization for a quench from the ground state of $H(0,0,0.23)$ and time evolved with $H(0.2,0.4,0.3)$ using mean-field theory, the second Born approximation and a TEBD calculation with maximum bond dimension $\chi=800$ and $L=200$.}
    \label{fig:2BADecay_GSQuench}
\end{figure}
In Fig.~\ref{fig:2BADecay_GSQuench} we show results for quench initialized in the ground state of $H(0,0,0.23)$ and time evolved with $H(0.2,0.4,0.3)$. 
Here the post-quench energy per site is $\epsilon_{\rm Quench}\approx 0.040$ corresponding to a mean-free path $\ell\approx 29$, which is much smaller than our system size of $L=200$. We observe that the second Born approximation clearly shows the decay of the oscillatory behaviour of the staggered magnetization. We plot the TEBD results only up to times $Jt=40$, where our criterion is agreement of the numerical results for bond dimensions $\chi=600$ and our maximal bond dimension $\chi_{\rm max}=800$. The TEBD data show the beginning of a decay, consistent with the second Born results.
In Figs~\ref{fig:2BADecay} and \ref{fig:2BADecay2} we show the behaviour of the staggered magnetization after quenches from thermal initial states.
\begin{figure}[ht]
    \centering
    \includegraphics[scale=0.5]{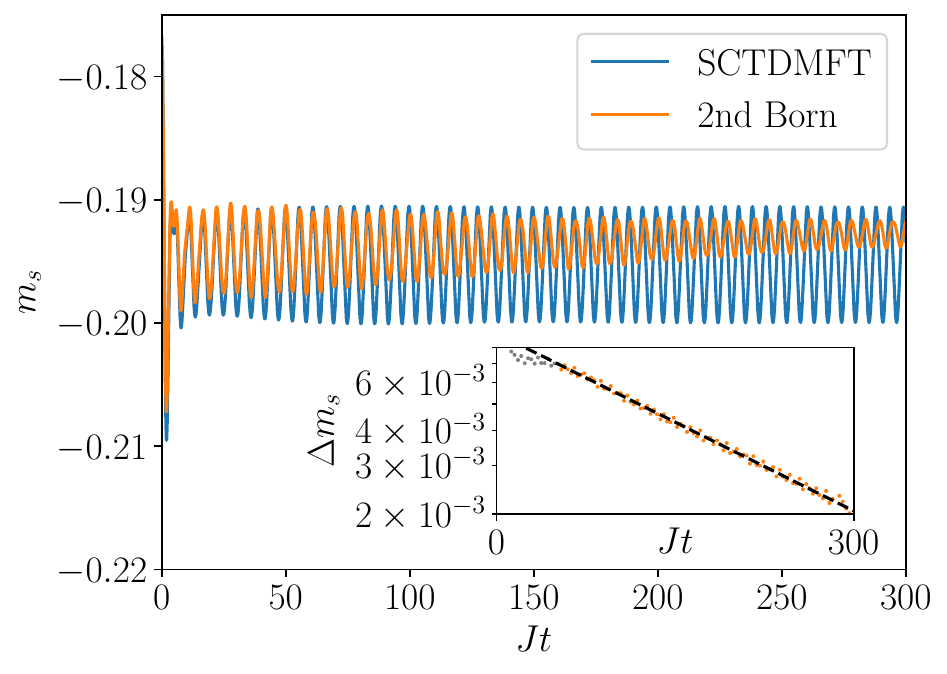}
    \caption{Main: Staggered magnetization for a quench from the thermal state $\rho(0,0.4,0.3,4.0)$ of the non-interacting system and time evolved with $H(0.2,0.4,0.3)$. $L=448$ used for the second Born approximation. The initial state is chosen such that the energy density is approximately the same as in Fig.~\ref{fig:2BADecay_GSQuench}. Inset: successive peak to peak amplitudes of the oscillations in the second Born approximation, with an exponential fit (dashed black line). The grey scatter points are excluded from the fit.}
      \label{fig:2BADecay}
\end{figure}
In Fig.~\ref{fig:2BADecay} we initialize the system in the thermal state of the non-interacting system at inverse temperature $J\beta_i=4$, which corresponds to the same energy density as in Fig.~\ref{fig:2BADecay_GSQuench}. We again observe decaying oscillations. In the inset in that figure, we estimate the decay time by fitting a decaying exponential $A\exp(-t/\tau)$ to the successive peak to peak amplitudes - the resulting decay time for this particular quench is $J\tau \sim 200$.
Finally, in Fig.~\ref{fig:2BADecay2}, we consider a lower temperature $J\beta_i=10$, which 
corresponds to energy per site $\epsilon_{\rm Quench}\approx 0.0037$ and mean free path $\ell\approx 250$.
\begin{figure}[ht]
    \centering
    \includegraphics[scale=0.5]{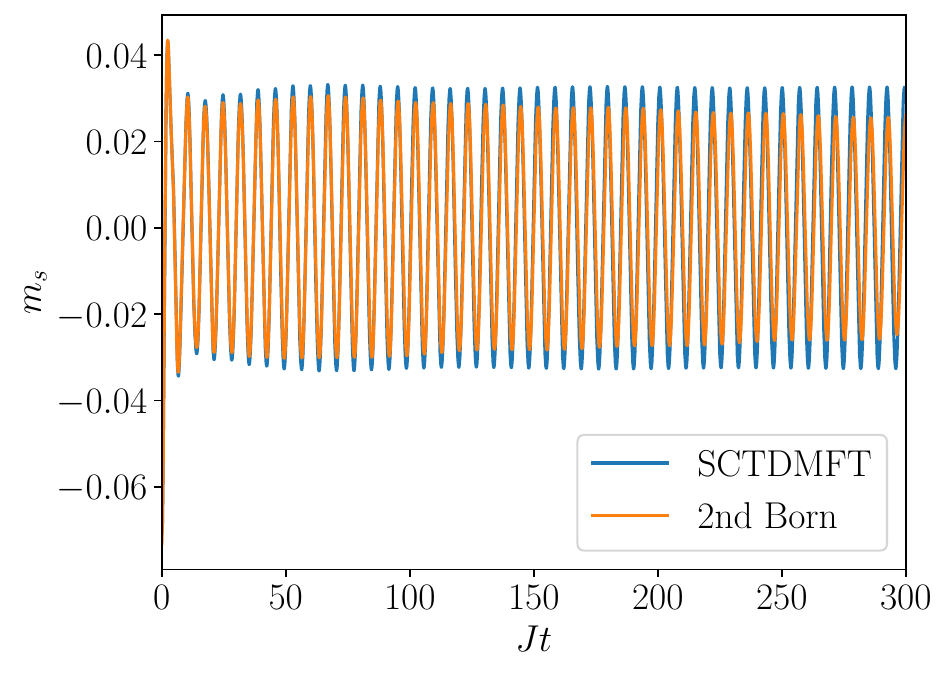}
    \caption{Staggered magnetization for a quench starting in the thermal state $\rho(0,0.4,0.10,10.0)$ and time evolved with $H(0.2,0.4,0)$ on a ring with $L=400$. Other than the finite pre-quench temperature $J\beta_i=10.0$ this is the same as Fig.~\ref{fig:TEBD_NN}.}
    \label{fig:2BADecay2}
\end{figure}
Here the oscillations are seen to decay very slowly.

%%%%%%%%%%%%%%%%%%%%%%%%%%%%%%%%%%%%%%%
\section{Summary and conclusions}
%%%%%%%%%%%%%%%%%%%%%%%%%%%%%%%%%%%%%%%
In this work we have carried out a detailed study of a mechanism that gives rise to long-lived oscillations in the expectation values of local observables after quantum quenches. This mechanism is very different from quantum scars and occurs after small quenches in interacting many-particle systems with an excitation gap, which generate a regime that can be understood in terms of a low-density gas of (long-lived) kinematically protected quasiparticles. Long-lived oscillations can then occur in expectation values of observables that have matrix elements between the ground state and excited states that contain a single quasiparticle. 

We have presented very strong evidence using a combination of matrix-product state methods and perturbative approaches based on truncations of the BBGKY hierarchy that these oscillations decay at late times in all models we have considered. This is an important difference to models with exact quantum scars \cite{moudgalya2018entanglement,Moudgalya2022Quantum}.

Our results show that the linear response theory prediction is upheld only at the level of self-consistent mean-field theory. Going beyond mean-field theory to the second Born approximation provides evidence of damping. For small interaction strengths $U$ the timescale of the decay is therefore generally expected to be $\mathcal{O}(U^{-2})$. Whilst it is not impossible that higher order corrections would cause the oscillations to remain at late times, this is highly unlikely as there is no reason to anticipate such an effect. Instead, truncating the BBGKY hierarchy at higher orders should merely modify the lifetime. We presented non-perturbative numerics using TEBD for the oscillations in both the spin-1 chain (Fig.~\ref{fig:Spin1-MediumQuench}) and dimerized XXZ chain (Fig.~\ref{fig:2BADecay_GSQuench}) which indicate our qualitative conclusions that the oscillations damp at intermediate times are robust to including higher orders.

Our results differ from the prediction made in \cite{Delfino2020Persistent, Delfino2022persistent} that the oscillations have infinite lifetime regardless of the quench strength $\lambda$. Whilst those papers are formulated in the continuum, the same arguments made therein lead to oscillations on the lattice that we have shown to decay. 

Our perturbative analysis generalizes straightforwardly to other interacting fermion and boson models with interactions that are local with regards to the elementary excitations of the unperturbed theory. We note that the much-studied Ising chain in a tilted field \cite{Banuls2011Strong,KormosRealtime2016,Lin2017Quasiparticle,Collura2018Dynamical,Hodsagi2018Quench,Robinson2019Signatures,Scopa_2022,Birnkammer_2022} does not fall within this class of models. This is because when viewed as a perturbation of the transverse-field Ising chain, which maps onto non-interacting fermions by the Jordan-Wigner transformation, the perturbing operator is not local in terms of these fermions, i.e. involves interaction vertices with arbitrarily large numbers of particles. This precludes employing approaches based on truncating the BBGKY hierarchy for the fermionic degrees of freedom. 
%However, whilst the methods we use to study the decay are not applicable to this case, it is still physically expected that local observables should relax to a stationary value.

%%%%%%%%%%%%%%%%%%%%%%%%%%
\begin{acknowledgments}
This work was supported by the EPSRC under grant EP/S020527/1. We are very grateful to Bruno Bertini for providing us with the second Born approximation code used in Refs~\onlinecite{Bertini15Prethermalization,Bertini16Thermalization}. 
\end{acknowledgments}

\appendix

\section{Free limit}
\label{App:Delta0}
In this appendix we briefly discuss the properties of the model \fr{Eq:HSpin} and its fermionic  version \fr{Eq:HFerm} that are important to understanding the main text at its free point $\Delta=0$. The spins $S^\alpha_m$ obey periodic boundary conditions $S^\alpha_{m+L}=S^\alpha_{m}$. Consequently the fermion operators obey boundary conditions
\begin{equation}
    c_{n+L}=-(-1)^{\hat{N}} c_n \ , 
\end{equation}
where $\hat{N}=\sum_n c_n^\dag c_n$ is the total fermion number. At half filling we must therefore consider periodic (Ramond) boundary conditions when there are an odd number of unit cells and anti-periodic (Neveu-Schwarz) boundary conditions when there are an even number.

\begin{figure}[t!]
    \centering
    \includegraphics[scale=0.7]{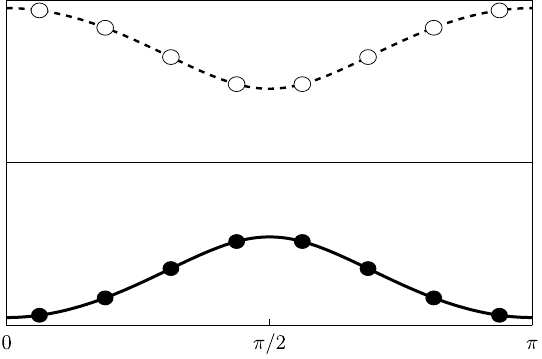}
    \caption{Dispersion at $\Delta=0$, with the ground state indicated. Filled circles indicate states that are occupied and empty circles ones that are unoccupied, drawn at $L=16$ for clarity. Note that $k=\pi/2$ is not an allowed single particle state for any finite system size.}
    \label{fig:FermiSea}
\end{figure}

The presence of the staggering causes the states at $k$ and $k+\pi$ to hybridise into two states which we denote $(k,+)$ and $(k,-)$ which have a dispersion relation that can be found from Eq.~(\ref{Eq:HMFT_BlockDiag}) upon setting $\Delta=0$ and diagonalizing the $2\times 2$ matrix $\Tilde{h}_{\mu\nu}$. The result is that the free part of the Hamiltonian is diagonalised with the canonical transformation:
\begin{equation}
    c_{m}=\sqrt{\frac{2}{L}}\sum_{k,\mu} e^{-ikm}\gamma_{m,\mu}(k) b_\mu(k) \ , \label{Eq:CanonicalTransform1}
\end{equation}
where the Bogoliubov co-efficients $\gamma_{\mu}(k)$ are
\begin{align}
   &\gamma_{2m,0} = -e^{-i\varphi_k}\sin\frac{\theta_k}{2} \ , 
   \gamma_{2m-1,0} = \cos\frac{\theta_k}{2} \ , \nn
   & \ \gamma_{2m,1} = e^{-i\varphi_k} \cos\frac{\theta_k}{2} \ , 
    \ \gamma_{2m-1,1} = \sin\frac{\theta_k}{2} \ , \label{Eq:CanonicalTransform2}
\end{align}
here the two Bogoliubov angles $\varphi_k$ and $\theta_k$ are given by
\begin{align}
&\cos \frac{\theta_k}{2} = \sqrt{\frac{\epsilon_k + h_s}{2\epsilon_k}} \ , \ \sin \frac{\theta_k}{2} = \sqrt{\frac{\epsilon_k - h_s}{2\epsilon_k}} \ , \nn 
&e^{-i \varphi_k(\alpha)} = \frac{-\cos k+ i \alpha\sin k}{\sqrt{\cos^2 k+ \alpha^2\sin^2k}} \ , \label{Eq:CanonicalTransform3}
\end{align}
where $\epsilon_k$ is the dispersion of the free part of the Hamiltonian and equal to
\begin{equation}
    \epsilon_\pm(k) = \sqrt{\alpha^2 J^2+h_s^2+(1-\alpha^2)J^2\cos^2 k} \ . \label{Eq:FreeDisp}
\end{equation}
The ground state is then a Fermi sea where the entire $(-)$ band is filled and the $(+)$ band is empty, illustrated in Figure \ref{fig:FermiSea}.
This picture of the ground state at $\Delta=0$ allows us to understand why the ED results are poorly converged with respect to system size when estimating the gap for small $\Delta$. In the thermodynamic limit the minimum energy excitation corresponds to a hole in the filled band at $k=\pi/2$ and a particle in the empty upper band at $k=\pi/2$, with a corresponding gap of $E_{\rm gap}(\infty)=2\epsilon_+(\pi/2)=2\sqrt{\alpha^2J^2+h_s^2}$. However, if the number of unit cells $L/2$ is even then we must work in the Neveu-Schwarz sector and so have anti-periodic boundary conditions with $k=(2n+1)\pi/L$, which never equals exactly $\pi/2$. Likewise for an odd number of unit cells we work in the Ramond sector where $k=2n\pi/L\neq \pi/2$. 
For finite $L$ the gap is therefore
\begin{equation}
 E_{\rm gap}(L) = 2\epsilon_+\left(\frac{\pi}{2}+\frac{\pi}{L}\right)  = E_{\rm gap}(\infty) + \delta E_{\rm gap}(L) \ , 
\end{equation}
for large $L$ the finite size effects become
\begin{align}
\delta E_{\rm gap}(L)=&\odiff{^2\epsilon}{k^2}\Big|_{k=\pi/2}\frac{\pi^2}{L^{2}}=J\frac{1-\alpha^2}{\sqrt{\alpha^2+h_s^2}}\frac{\pi^2}{L^{2}} \ , 
\end{align}
These considerations motivate the following form of a fitting function 
\begin{equation}
    E_{\rm gap}(L)=E_{\rm gap}(\infty)+BL^{-2}+ CL^{-4} \ . \label{Eq:EDFits}
\end{equation}
This functional form indeed provides a good description of our numerical results for the gap deduced from ED on $L\in\{20,22,24,26,28,30\}$ sites when $\Delta$ is small.

\section{Equations of motion in Second Born approximation}
\label{App:2BA}

The method we use for deriving the equations of motion \fr{Eq:2BA_EoM} is given in more detail in \cite{Bertini15Prethermalization,Bertini16Thermalization}, here we simply briefly recap the main points for completeness. The first step is to diagonalize the free part of the Hamiltonian using Eqs.~\fr{Eq:CanonicalTransform1}-\fr{Eq:CanonicalTransform3}.

The next step is to rewrite the interaction in this basis. Whilst in the original real-space basis the interaction is the same as that considered in \cite{Bertini15Prethermalization,Bertini16Thermalization}, we find a different final form since we obtain different $b_\mu(k)$ of the free part of the Hamiltonian, which agrees with the expression there when setting $h_s=0$. We antisymmetrize the interaction in the first and second pairs of indices
\begin{equation}
V_{\bm\mu}({\bm k}) = - \frac{1}{4} \sum_{P\in \Z_2\times \Z_2} \sgn(P) V'_{P({\bm \mu})}(P({\bm k}))~.
\end{equation}
Here $P$ is an element of $\Z_2\times \Z_2$ where the first $\Z_2$ swaps $\mu_1\leftrightarrow \mu_2, k_1 \leftrightarrow k_2$ and the second $\Z_2$ factor acts likewise on $3,4$. We define $\sgn(P)$ as the product of the sign of each permutation. The unsymmetrized interaction components are equal to
\begin{widetext}
\begin{align}
V'_{\vec{\mu}}(\vec{k}) = \frac{2e^{i(k_2-k_3)}}{L}  \bigg\lbrack&  g_{\mu_1}(k_1) f_{\mu_2}(k_2) f_{\mu_3}(k_3) g_{\mu_4}(k_4) e^{i(\varphi_{k_1}-\varphi_{k_4})}
\big(\delta_{k_1+k_2,k_3+k_4}+\delta_{k_1+k_2-k_3-k_4,\pm \pi}\big)\nn
& + f_{\mu_1}(k_1) g_{\mu_2}(k_2) g_{\mu_3}(k_3) f_{\mu_4}(k_4) e^{i(\varphi_{k_2}-\varphi_{k_3})} \big(\delta_{k_1+k_2,k_3+k_4}-\delta_{k_1+k_2-k_3-k_4,\pm \pi}\big) \bigg\rbrack\ ,
        \label{Eq:VInt}
    \end{align}
\end{widetext}
where we have defined
\begin{eqnarray}
    f_\mu(k) &=& (1-\mu)\cos\frac{\theta_k}{2} + \mu \sin \frac{\theta_k}{2} \ , \nn 
    g_\mu(k) &=& \mu \cos \frac{\theta_k}{2} - (1-\mu) \sin \frac{\theta_k}{2} \ , 
\end{eqnarray}
The Heisenberg equations of motion are
\begin{eqnarray}
    \frac{\partial \hat{n}_{\mu\nu}(k,t)}{\partial t} = i\epsilon_{\mu\nu}\hat{n}_{\mu \nu} + i \Delta \sum Y_{\mu\nu}^{\vec{\mu}}(k,\vec{q})\hat{A}_{\vec{\mu}}(\vec{q}) \ , \label{Eq:Heisenberg}
\end{eqnarray}
where $\epsilon_{\mu\nu}(k)=\epsilon_\mu(k)-\epsilon_\nu(k)$ and the coefficients of the quartic operators appearing are given explicitly in terms of the interaction potential $V_{\vec{\mu}}(\vec{k})$ by
\begin{align}
    &Y_{\alpha\beta}^{\vec{\mu}}(k,\vec{q}) = \delta_{\beta\mu_4}\delta_{k,q_4}V_{\mu_1\mu_2 \mu_3 \alpha}(\vec{q})  
    +  \delta_{\beta\mu_3}\delta_{k,q_3}V_{\mu_1\mu_2 \alpha \mu_4 }(\vec{q}) \nn 
    &-  \delta_{\alpha\mu_2}\delta_{k,q_2}V_{\mu_1\beta \mu_3 \mu_4 }(\vec{q}) - \delta_{\alpha \mu_1}\delta_{k,q_1}V_{\beta\mu_2 \mu_3 \mu_4}(\vec{q}) \ . \ 
\end{align}

The quartic operators $\hat{A}_{\vec{\mu}}(\vec{k})$ likewise evolve according to the following Heisenberg equations of motion
\begin{align}
&\frac{\partial}{\partial t}\hat{A}_{\boldsymbol{\mu}}(\bm{k},t) =
 i {E}_{\boldsymbol{\mu}}(\bm{k})\hat{A}_{\boldsymbol{\mu}}(\bm{k},t) \label{Eq:Heisenberg4point} \\
&\qquad +i \Delta \sum_{\boldsymbol{\gamma}}\sum_{\bm{q}{>0}} {V}_{\boldsymbol{\gamma}}(\bm{q})
\left[\hat{A}_{\boldsymbol{\gamma}}(\bm{q},t),\hat{A}_{\boldsymbol{\mu}}(\bm{k},t)\right]~, \nonumber 
\end{align}
where $E_{\vec{\mu}}(\vec{k}) = \epsilon_{\mu_1}(k_1) + \epsilon_{\mu_2}(k_2) - \epsilon_{\mu_3}(k_3) - \epsilon_{\mu_4}(k_4)$. This equation contains products of up to six fermionic operators on the right hand side, and carrying on in this way generates the BBGKY hierarchy of equations of motion. 
The second Born approximation consists of truncating at this level which we do by formally integrating Eq.~\fr{Eq:Heisenberg4point} to obtain an integral expression for $\hat{A}_{\vec{\mu}}(\vec{k},t)$ in terms of its value at $t=0$ which can be substituted into the Heisenberg equations of motion \fr{Eq:Heisenberg}, which gives
\begin{align}
       & \frac{\partial}{\partial  t}{n}_{\mu\nu}(k,t)=i\epsilon_{\mu\nu}(k) n_{\mu\nu}(k,t) \label{Eq:EOM} \\ 
        &+i \Delta \sum_{{\vec \eta}}\sum_{{\vec{q}{>0}}}{Y}_{\mu\nu}^{\boldsymbol{\eta}}(k,\vec{q})\braket{\hat{A}_{\boldsymbol{\eta}}(\vec{q},0)} e^{i  t {E}_{\boldsymbol{\eta}}(\vec{q})} \nn
        &+\Delta^2\sum_{\boldsymbol{\eta},\boldsymbol{\gamma}}\sum_{\vec{q},\vec{p}{>0}}  \int_{0}^{t}\!{\rm d}s~\braket{\hat{A}_{\boldsymbol{\gamma}}(\vec{p},s)\hat{A}_{\boldsymbol{\eta}}(\vec{q},s)} \times \nn 
        &\left({Y}_{\mu\nu}^{\boldsymbol{\gamma}}(k,\vec{p}) 
        e^{i  (t-s){E}_{\boldsymbol{\gamma}}(\vec{p})}{V}_{\boldsymbol{\eta}}(\vec{q}) 
       - (\vec{p},\boldsymbol{\gamma})\leftrightarrow (\vec{q},\boldsymbol{\gamma}) \right) \nonumber \ . 
\end{align}
Neglecting the 4- and 6-particle cumulants in \fr{Eq:EOM} then leads to Eq.~\fr{Eq:2BA_EoM}.

%\end{widetext}

%\bibliography{references}% Produces the bibliography via BibTeX.

%apsrev4-2.bst 2019-01-14 (MD) hand-edited version of apsrev4-1.bst
%Control: key (0)
%Control: author (72) initials jnrlst
%Control: editor formatted (1) identically to author
%Control: production of article title (-1) disabled
%Control: page (0) single
%Control: year (1) truncated
%Control: production of eprint (0) enabled
%

\end{document}